\documentclass[conference]{IEEEtran}
\IEEEoverridecommandlockouts
\usepackage{cite}
\usepackage{amsmath,amssymb,amsfonts}
\usepackage{graphicx}
\usepackage{textcomp}
\usepackage{xcolor}
\usepackage{url}

\usepackage[normalem]{ulem}
\useunder{\uline}{\ul}{}

\usepackage{algorithm}
\usepackage[noend]{algpseudocode}

\algrenewcommand\alglinenumber[1]{#1}
\algrenewcommand\algorithmicindent{1em}

\usepackage{amsthm}
\newtheoremstyle{def_style}
  {}          
  {}          
  {}          
  {}          
  {\bfseries} 
  {.}         
  {.5em}      
  {}          

\theoremstyle{def_style}

\theoremstyle{def_style}
\newtheorem{prob}{Problem}

\usepackage{caption}
\usepackage{subcaption}

\DeclareMathOperator*{\argmin}{arg\,min}

\usepackage{enumitem}
\setlist[itemize]{leftmargin=*, topsep=0pt, itemsep=0pt, parsep=0pt, partopsep=0pt}
\setlist[enumerate]{leftmargin=*, topsep=0pt, itemsep=0pt, parsep=0pt, partopsep=0pt}

\usepackage{multirow}

\begin{document}

\title{Time Series Synthesis Using\\the Matrix Profile for Anonymization}

\author{
\IEEEauthorblockN{Audrey Der$^{1,2}$, Chin-Chia Michael Yeh$^2$, Yan Zheng$^2$, Junpeng Wang$^2$, Huiyuan Chen$^2$, \\ Zhongfang Zhuang$^2$, Liang Wang$^2$, Wei Zhang$^2$, Eamonn Keogh$^1$}
\IEEEauthorblockA{$^1$\textit{Computer Science and Engineering at University of California, Riverside}, $^2$\textit{Visa Research} \\
ader003@ucr.edu, \{miyeh, yazheng, junpenwa, hchen, zzhuang, liawang, wzhan\}@visa.com, eamonn@cs.ucr.edu}
}


\maketitle

\begin{abstract}
Publishing and sharing data is crucial for the data mining community, allowing collaboration and driving open innovation.
However, many researchers cannot release their data due to privacy regulations or fear of leaking confidential business information.
To alleviate such issues, we propose the \textit{Time Series Synthesis Using the Matrix Profile} (TSSUMP) method, where synthesized time series can be released in lieu of the original data.
The TSSUMP method synthesizes time series by preserving similarity join information (i.e., Matrix Profile) while reducing the correlation between the synthesized and the original time series.
As a result, neither the values for the individual time steps nor the local patterns (or \textit{shapes}) from the original data can be recovered, yet the resulting data can be used for downstream tasks that data analysts are interested in.
We concentrate on similarity joins because they are one of the most widely applied time series data mining routines across different data mining tasks.
We test our method on a case study of ECG and gender masking prediction. 
In this case study, the gender information is not only removed from the synthesized time series, but the synthesized time series also preserves enough information from the original time series. 
As a result, unmodified data mining tools can obtain near-identical performance on the synthesized time series as on the original time series.
\end{abstract}




\section{Introduction}\label{sec:introduction}
When releasing datasets to the public, the most common concerns are privacy-related issues. 
This is because preserving the privacy of the data donors is one of the most important ethical and social responsibilities for data mining researchers. 
Moreover, researchers must typically adhere to data protection and privacy laws like HIPAA~\cite{hippa}, GDPR~\cite{gdpr}, and PIPA~\cite{pipa}. 
As a result, the dataset must be properly anonymized before releasing it to the public. 
We recognize that ``privacy" is a term with precedent and is well-defined in other domains; any of our references to ``anonymization" are informal unless otherwise noted.
In this work, our focus is on the anonymization of time series through  solving the \textit{time series substitution problem}:

\begin{prob}
\label{prob:substitution}
    Let the function~$c(\cdot, \cdot)$ compute the correlation between two inputs, and the set~$\mathbf{D}$ be a collection of time series data mining methods for different data mining tasks. 
    Given a time series~$T$, we want to define an anonymization function~$f(\cdot)$ which synthesis time series $\hat{T}=f(T)$ such that $c(T, \hat{T}) \approx 0$ and $d(T) \approx d(\hat{T})\,\forall\,d \in \mathbf{D}$.
\end{prob}

\noindent The goal of Problem~\ref{prob:substitution} is to synthesis a time series~$\hat{T}$ from the original time series $T$. 
The synthesis function~$f(\cdot)$ in Problem~\ref{prob:substitution} needs to satisfy two conditions: 1) $c(T, \hat{T}) \approx 0$ and 2) $d(T) \approx d(\hat{T})\,\forall\,d \in \mathbf{D}$.
The role of the first condition is to ensure~$T$ and $\hat{T}$ have different local patterns; we call it the \textit{local pattern condition}. 
It is crucial to conceal the local patterns because, as demonstrated in our experiment (Section~\ref{sec:case_ecg}), they have the potential to reveal traits about the individuals who generate the time series.
We use Pearson correlation~$c(\cdot, \cdot)$ to ensure $T$ and $\hat{T}$ are not linearly correlated with each other.
This safeguards against malicious attackers trying to recover $T$ from $\hat{T}$ as the solution to $\argmin_{\hat{T}}\textsc{PearsonCorr}(T, \hat{T})^2$ is not unique.

The goal for the second condition is to maintain the utility of~$T$.
We hypothesize we could satisfy $d(T) \approx d(\hat{T})\,\forall\,d \in \mathbf{D}$ by enforcing $\textsc{MatrixProfile}(T) \approx \textsc{MatrixProfile}(\hat{T})$.
The \textit{Matrix Profile} (MP) is an effective method to summarize the distance between subsequences and their nearest neighbors~\cite{yeh2016matrix,yeh2018time}. 
We choose to preserve such structure information when generating~$\hat{T}$ because  there is an increasing realization that such information is sufficient for many time series data mining tasks. 
These include segmentation, rule discovery, time series joins, and anomaly detection~\cite{mp_website}. 
By preserving the MP from $T$ to $\hat{T}$, we are able to maintain sufficient information to ensure success in some $d \in \mathbf{D}$. 
This condition, which focus on the nearest neighbor relationship between subsequences within a time series (a type of global structure), is referred to as the \textit{global structure condition}.

We note the choice of Pearson correlation and MP is only \textit{one} of many possible formulations for the time series substitution problem.
This is the initial attempt to solve this problem, and we limit our scope to this particular variant of the problem for proof-of-concept purposes. 
We defer the investigation of other variations for future work.

\begin{figure*}[t]
\centerline{
\includegraphics[width=0.8\linewidth]{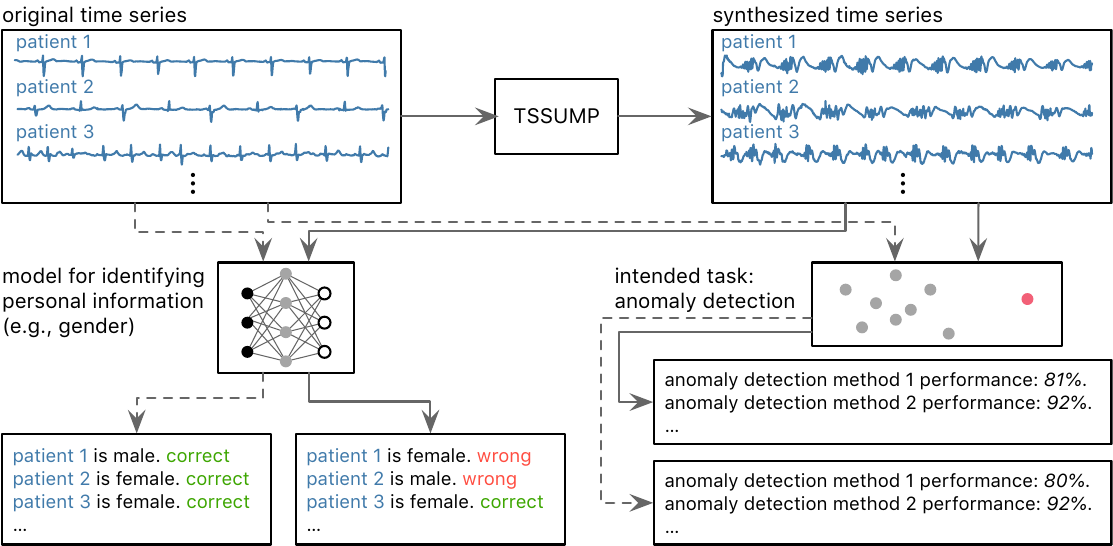}
}
\caption{
A process diagram of using TSSUMP to release electrocardiogram (ECG) time series. Models trained on existing public ECG datasets to identify sensitive information about patients (e.g., gender) are unable to do so while researchers can still use existing time series analysis methods for tasks, such as anomaly detection.
More details can be found in Section~\ref{sec:case_ecg}. 
}
\label{fig:motivation}
\end{figure*}

We use the case study in Fig.~\ref{fig:motivation} to illustrate how the TSSUMP method can anonymize time series by addressing Problem \ref{prob:substitution} on a real-world dataset.
Suppose an electrocardiogram (ECG) time series anomaly detection (TSAD) benchmark is made available in the public domain, comprised comprised of data collected from multiple patients, and does not include any accompanying metadata that contains personal information (e.g. gender, age, pre-existing health conditions).

However, an individual might build a model capable of identifying, in whole or part, sensitive information from this ECG TSAD benchmark using \textit{other} ECG datasets in the public domain~\cite{goldberger2000physiobank}.
This potential information recovery is feasible because the \textit{local patterns} present in $T$ (such as individual heartbeats) exhibit similarity to the patterns observed in other ECG datasets.
As depicted in Fig.~\ref{fig:motivation}, a malicious individual could compromise the privacy of patients' data by utilizing a gender prediction model trained on datasets from~\cite{goldberger2000physiobank} to deduce the gender of each patient.

Using the TSSUMP method, the synthesized time series contains different \textit{local pattern} while the \textit{utility} of the original time series is maintained. 
If we release the synthesized time series in lieu of the original ECG time series, the malicious individual's gender prediction model will degrade to the default rate when classifying patients in our new benchmark.
This deterioration arises from the discrepancy between the shape of the heartbeat in the new synthesized benchmark and the patterns observed in~\cite{goldberger2000physiobank}. 
By preserving the utility, the synthesized dataset ensures consistent performance across various anomaly detection methods. 
Thus, it can serve as a suitable replacement for the original ECG dataset, enabling researchers to develop methods for time series tasks.

Our contributions include:
\begin{itemize}
\item We define the time series substitution problem as a means to address the anonymization problem for time series data. 
\item We propose an initial successful solution leveraging the MP to preserve the utility of original time time series while the local patterns are unrecoverable from $\hat{T}$.
\item We demonstrate the effectiveness of TSSUMP using a real-world ECG dataset.
\end{itemize}

\section{Methodology}
The synthesis process of TSSUMP involves solving an optimization problem.
In this section, we focus on discussing the loss function utilized in TSSUMP.
For more details and the implementation of the optimization algorithm, please refer to our project website~\cite{supplementwebsite}.

We design the loss function~$\mathcal{L}$ so that the solution of $\argmin_{\hat{T}} \mathcal{L}(T, \hat{T})$ satisfies the local pattern and global structure conditions, and it is shown in Eq.~\ref{eq:loss}.
\begin{equation}
\label{eq:loss}
    \mathcal{L} = \mathcal{L}_{\text{local}} + \mathcal{L}_{\text{global}}
\end{equation}
where the $\mathcal{L}_{\text{local}}$ is the local pattern condition loss, $\mathcal{L}_{\text{global}}$ is the global structure condition loss for preserving utility.

\begin{figure}[ht]
    \centering
    \includegraphics[width=0.99\linewidth]{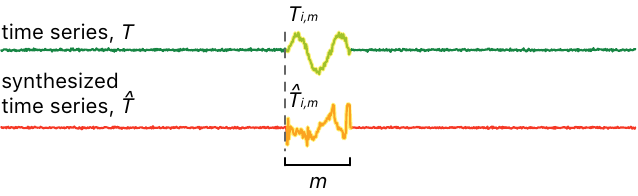}
    \caption{\sloppy
    The local pattern condition loss~$\mathcal{L}_{\text{local}}$ is computed with $\textsc{PearsonCorr}(T_{i,m},\hat{T}_{i,m})^2$ using subsequences from the same temporal location (i.e., $i$th time stamp).}
    \label{fig:loss_corr}
\end{figure}

Since~$\mathcal{L}_{\text{local}}$ focuses on the local patterns (or \textit{subsequences}), as the name suggests, $\mathcal{L}_{\text{local}}$ is defined as:
\begin{equation}
\label{eq:pri_loss}
\mathcal{L}_{\text{local}} = \textsc{PearsonCorr}(T_{i,m}, \hat{T}_{i,m})^2
\end{equation}
where $T_{i,m}$ is the $i$th subsequence in $T$, $\hat{T}_{i,m}$ is the $i$th subsequence in $\hat{T}$, and $m$ is the subsequence length.
In the example shown in Fig.~\ref{fig:loss_corr}, the Pearson correlation is computed between the subsequence~$T_{i,m}$ from the original time series~$T$ and the subsequence~$\hat{T}_{i,m}$ from the synthesized time series~$\hat{T}$.
The subsequences are both the $i$th subsequence from their corresponding time series.
    
\begin{figure}[ht]
    \centering
    \includegraphics[width=0.99\linewidth]{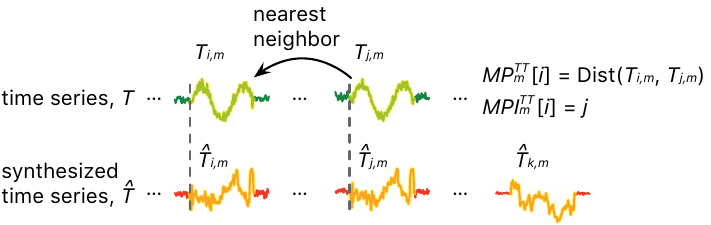}
    \caption{\sloppy
    Given the sampled subsequence index~$i$, $\mathcal{L}_{\text{distance}}$ and $\mathcal{L}_{\text{identity}}$ are computed using the subsequences highlighted in the figure (see Eq. \ref{eq:dist_loss} and \ref{eq:id_loss}).
    }
    \label{fig:loss_mp}
\end{figure}

The~$\mathcal{L}_{\text{global}}$ loss term preserves the MP and MP Index (MPI) of $T$ when synthesizing $\hat{T}$.
Since the MP and MPI store the distance and identity of the nearest neighbor for each subsequence, $\mathcal{L}_{\text{global}}$ is defined as:
\begin{equation}
    \label{eq:utility_loss} \footnotesize
    \mathcal{L}_{\text{global}} = \mathcal{L}_{\text{distance}} + \mathcal{L}_{\text{identity}}
\end{equation}
where~$\mathcal{L}_{\text{distance}}$ is the distance loss and~$\mathcal{L}_{\text{identity}}$ is the identity loss for retaining the distance and identity information, respectively.
If $T_{j,m}$ is $T_{i,m}$'s nearest neighbor in $T$, the distance loss~$\mathcal{L}_{\text{distance}}$ is defined as:
\begin{equation}
\label{eq:dist_loss} \footnotesize
\mathcal{L}_{\text{distance}} = \left( \textsc{Dist}(T_{i,m}, T_{j,m}) - \textsc{Dist}(\hat{T}_{i,m}, \hat{T}_{j,m})\right)^2
\end{equation}
where $\textsc{Dist}(\cdot,\cdot)$ is a function that computes the $z$-normalized Euclidean distance between the inputs.
The distance loss~$\mathcal{L}_{\text{distance}}$ ensures the distance between $T_{i,m}$ and its  nearest neighbor~$T_{j,m}$ equal to the distance between their counterparts in $\hat{T}$.
If $\mathcal{L}_{\text{distance}}$ is minimized, $\hat{T}$ and $T$ will have similar MPs.
Fig.~\ref{fig:loss_mp} shows the relationship between $T_{i,m}$, $T_{j,m}$, $\hat{T}_{i,m}$, and $\hat{T}_{j,m}$.

To preserve the nearest neighbor identity information, we need to consider the distances between a subsequence~$\hat{T}_{i,m}$ and the other subsequences~$\hat{T}_{k,m}$ in~$\hat{T}$.
If the nearest neighbor of $T_{i,m}$ is $T_{j,m}$, $\textsc{Dist}(\hat{T}_{i,m},\hat{T}_{j,m})$ needs to be smaller than $\textsc{Dist}(\hat{T}_{i,m},\hat{T}_{k,m})$ for any $k \in \{0,\cdots,n-m+1\}\setminus\{i,j\}$.
To enforce such a relationship, we have defined the identity loss~$\mathcal{L}_{\text{identity}}$ as:
\begin{equation}
\label{eq:id_loss} \footnotesize
\mathcal{L}_{\text{identity}} = \textsc{ReLU} \left( \textsc{Dist}(\hat{T}_{i,m}, \hat{T}_{j,m}) - \textsc{Dist}(\hat{T}_{i,m}, \hat{T}_{k,m})\right)
\end{equation}
The term $\textsc{Dist}(\hat{T}_{i,m},\hat{T}_{j,m})$ computes the distance between a supposed nearest neighbor pair of subsequences, and the term $\textsc{Dist}(\hat{T}_{i,m}, \hat{T}_{k,m})$ computes the distance between a supposed non-nearest neighbor pair.
Because we want to ensure that $\textsc{Dist}(\hat{T}_{i,m},\hat{T}_{j,m}) < \textsc{Dist}(\hat{T}_{i,m},\hat{T}_{k,m})$, Eq.~\ref{eq:id_loss} will only result in non-zero value if such relationship is violated.
In other words, minimizing Eq.~\ref{eq:id_loss} ensures that the relationship holds in time series~$\hat{T}$.
Fig.~\ref{fig:loss_mp} shows the relationship between the subsequences used for computing $\mathcal{L}_{\text{identity}}$.

\section{Experiment}
\label{sec:case_ecg}
We use the ECG time series from MIT-BIH Long-Term ECG Database~\cite{goldberger2000physiobank} to validate the claims we presented in Fig.~\ref{fig:motivation}, where the targeted data mining task is time series anomaly detection (TSAD).
Recall that our claims are 1) that it is possible to build a classifier to detect personal information from heartbeat, 2) TSSUMP can reduce the risk of uncovering personal information from synthesized time series, and 3) the synthesized time series produced by TSSUMP is a suitable proxy for the original in TSAD benchmarks. 

There are seven patients in the ECG database.
To validate claims \#1 and \#2, we use 1,000 randomly extracted samples from each patient's \textit{first} 100,000 data points to build a gender/age classifier with Support Vector Machine (SVM)~\cite{pedregosa2011scikit,chang2011libsvm} using default \cite{pedregosa2011scikit} parameters.
The length of each sample is 100, and a total of 7,000 training samples (7 patients $\times$ 1,000 samples) are extracted from the database.
The gender and age classifiers are trained as binary classifiers, with the gender classes as male/female and the age classes as under/over 60 years old. The SVM models are relatively easy to train with off the shelf tools~\cite{pedregosa2011scikit,chang2011libsvm} compared to deep learning models~\cite{paszke2019pytorch}.

\subsubsection*{Validating Claims 1 and 2}
We randomly extracted 1,000 samples from the \textit{second} 100,000 data points of each patient.
We extract 7,000 test samples from both the original time series and the time series generated by TSSUMP. 
To ensure the comparison is fair, the samples are extracted from the same temporal location for both original and synthesized series.

As both classifiers are built for binary problems, we use the F1-score as the performance measurement.
When the classifier is applied to the test samples from \textit{original} time series, the F1-score for gender is 0.8585 and the F1-score for age is 0.9276.
In other words, even if the gender/age information is not included in the released dataset, a malicious individual could likely uncover such information using models build from public domain data.
However, when we apply the classifier on the \textit{substitute} time series, the F1-score drops to 0.4635 for gender and 0.3618 for age, indicating that this personal information is protected by TSSUMP. 

\subsubsection*{Validating Claim 3}
We conduct TSAD experiments similar to~\cite{yeh2022error}.
We extract the \textit{second} 100,000 data points of each patient for the experiment, using the normal heartbeats from the first 50,000 data points as the training data and the second 50,000 data points as the test data.
We apply multiple TSAD algorithms to both the original and the synthetic time series. 
The TSAD algorithms tested in our experimentation are shown in Table~\ref{tab:tsad_algs}.

\begin{table}[ht]
    \centering
    \caption{The tested TSAD algorithms. Vector space and subsequence based methods are denoted with an asterisk(*).} 
    \label{tab:tsad_algs}
    \footnotesize
    \begin{tabular}{p{0.25\linewidth} || p{0.6\linewidth}}
    Library & Method \\ \hline \hline
    Scikit-learn~\cite{pedregosa2011scikit} & Local Outlier Factor (LOF)*, Isolation Forest (IForest)*, One-Class Support Vector Machine (OCSVM)*,  Autoregressive Model (Autoreg)* \\ \hline
    PyOD~\cite{zhao2019pyod} & Auto-Encoder (AE)*, Variational Auto-Encoder (VAE)*, Deep Support Vector Data Description (DeepSVDD)*, Anomaly Detection with Generative Adversarial Networks (AnoGAN)* \\ \hline
    TODS~\cite{lai2021tods} & Recurrent Neural Network (RNN) with Long Short-Term Memory (LSTM), RNN with Gated Recurrent Units (GRU),  Deep Autoencoding Gaussian Mixture Model (DAGMM)* \\ \hline
    SCAMP~\cite{zimmerman2019matrix} & Matrix Profile (MP)*
    \end{tabular}
\end{table}

The vectors for vector space methods are extracted using a sliding window with step a size of one, and vector/subsequence lengths are set to 100. For the other hyperparameters, we use the default setting from the implementation.

We use the \textit{Area Under the Receiver Operating Characteristic Curve} (AUC) to measure the quality of the anomaly score when compared to the ground truth labels.
We choose to use AUC instead of the F1-score because anomaly scores returned by the evaluated methods need to be converted to binary predictions to evaluate F1-scores.
As pointed out by Kim et al.~\cite{kim2021towards}, the binarization process for TSAD is itself a challenging problem.
Therefore, we use AUC to avoid the additional complication in converting the anomaly scores to binary predictions. 
The performance (averaged AUC over 7 patients) of the 12 methods on both original and synthetic time series is shown in Table~\ref{tab:ecg_ad}. 

\begin{table}[ht]
    \centering
    \caption{Comparison of different anomaly detection methods on original and synthesized time series.}
    \footnotesize
    \label{tab:ecg_ad}
    \begin{tabular}{l|c|ccl|c|c}
     & Original & Synth. & ~& & Original & Synth. \\ \cline{1-3} \cline{5-7}
    LOF & 0.9531 & 0.9444 ~& & DeepSVDD & 0.6167 & 0.6228 \\
    IForest & 0.7272 & 0.6977 ~& & AnoGAN & 0.5773 & 0.4703\\
    OCSVM & 0.8006 & 0.8256 ~& & LSTM & 0.4859 & 0.4861\\
    AutoReg & 0.5114 & 0.4964 ~& & GRU & 0.4867 & 0.4857\\
    AE & 0.9383 & 0.8882 ~& & DAGMM & 0.6443 & 0.6438 \\
    VAE & 0.8771 & 0.8695 ~& & MP & 0.9294 & 0.9387 \\
    \end{tabular}
\end{table}

Aside from AnoGAN, the same TSAD method archives similar performance on both datasets. 
One possible reason for the inconsistency associated with AnoGAN is that the loss function used in AnoGAN is known to be unstable~\cite{arjovsky2017towards,arjovsky2017wasserstein}. 
Broadly speaking, the experiment results reported in Table~\ref{tab:ecg_ad} validate claim \#3.

\section{Related Work}\label{sec:relatedworks}
Shou et al.~\cite{shou2011supporting} proposed a time series anonymization method that utilizes the $(k, P$)-anonymity privacy model~\cite{shou2011supporting}, focusing on protecting individual time series within the grouped or aggregated time series, and it differs in its goals compared to TSSUMP.
Ruta et al.~\cite{ruta2015fast} propose a method to summarize and anonymize time series by providing a summarization of the time series while preserving the relevant shape information from the original series as much as possible. 
In contrast, TSSUMP is designed to remove shape information.
Wang et al.~\cite{wang2020part} proposed the PART-GAN model, an extension of the conditional generative adversarial network, to guarantee differential privacy for anonymizing time series classification datasets, addressing a different problem than Problem \ref{prob:substitution}.

To the best of our knowledge, the proposed TSSUMP method stands as the only time series anonymization technique that focusing on removing local pattern information while preserving the utility of the original time series. 
Notably, it accomplishes its objective without any dependence on accompanying label information and without requiring the original dataset to contain multiple time series. 

\subsection{The Importance of the MPI}
While we believe that our paper is completely self-contained, here we take the liberty of further explaining the importance of the location information contained within the Matrix Profile Index (MPI).

We want to reiterate and expand on the text in our paper, because a reader who is not very familiar with time series data mining might balk at the idea that the actual shape of the subsequences may not be important, but the location of the subsequence’s nearest neighbors is. Gharghabi et. al. showed in~\cite{gharghabi_MP8,gharghabi2019domain}, that for semantic segmentation we only need the location of the subsequence’s nearest neighbors. More generally, there seems to be an emerging realization that such information is sufficient for most time series data mining tasks.

For example, one of the main reasons why the community does motif discovery is to understand the temporal relationship between the motifs. Suppose we compare and contrast the motifs in the New York Taxi demand, and the motifs in New York temperature. An analyst may make the following observations:

\begin{itemize}
    \item For both datasets, the majority of the subsequence’s motifs are to the day before (or after). This is what forecasters call classic persistence, the assumption that the situation will be the same tomorrow as it was today.
    \item Uniquely for Taxi, some of the subsequence’s nearest neighbors will be seven days away. This is because of the vagaries of the typical five-day workweek, both Saturday and Sunday are different to typical weekdays, and each other.
    \item The timing of some motifs across the two different datasets are related. Unusually hot or cold weather does greatly affect taxi demand.
    \item There are a handful of days in the year when both Taxi and Temperature motifs strongly violate classic persistence, these are usually caused by blizzards. Such events are classic anomalies identified in the Numenta anomaly benchmark~\cite{ahmad2017unsupervised}.
    \item Finally, if the analyst has access to taxi demand from other cities, she may be able to see interesting differences. For example, Philadelphia, PA is only 80 miles from New York and has similar weather and culture. However, in New York alone, the subsequence’s representing the second Monday of October show much lower persistence. Why is that? It appears that to be caused by Columbus Day, which is celebrated by New York’s Italian-American community but is increasing ignored in the rest of the USA.
\end{itemize}

Note that all the analytics above (and much more) can be conducted with only access to knowledge of the location of the subsequence’s nearest neighbors.

\section{Conclusions}\label{sec:conclusion}
We introduced the \textit{Time Series Synthesis Using the Matrix Profile} (TSSUMP) method which takes an input time series~$T$ and synthesizes a new time series~$\hat{T}$ that preserves similarity join information.
The consideration of other sensitive characteristics (e.g., protection against location-based attacks~\cite{zhang2023pmp}), improvement of the utility of the synthesized series, enhancement of the scalability of the synthesizing process~\cite{yeh2022embedding, yeh2023sketching}, testing on subsequence classification problems~\cite{yeh2023egonetwork}, and evaluating privacy on a theoretical level are other areas of future work.
Our code and results are available in perpetuity at~\cite{supplementwebsite}.

\bibliographystyle{ieeetr}
\bibliography{section/reference}

\end{document}